\input harvmac.tex

\noblackbox
\lref\maxqm{M. Claudson and M.B. Halpern, Nucl. Phys. {\bf B250}
(1985) 689\semi
M. Baake, P. Reinicke, and V. Rittenberg, J. Math. Phys. {\bf 26} (1985)
1070\semi
R. Flume, Ann. Phys. {\bf 164} (1985) 189.} 
\lref\Sen{A. Sen, Phys. Rev. {\bf D54} (1996) 2964, hep-th/9510229\semi
A. Sen, Mod. Phys. Lett. {\bf A11} (1996) 827, hep-th/9512203.} 
\lref\ulfI{U. Danielsson, G. Ferretti, and B. Sundborg,
hep-th/9603081.}
\lref\KP{D. Kabat and P. Pouliot, {Phys. Rev. Lett.}
{\bf 77} (1996) 1004, hep-th/9603127.}
\lref\joenew{J. Polchinski, hep-th/9611050.}
\lref\ginsparg{P. Ginsparg, Phys. Rev. {\bf D35} (1987) 648.}
\lref\nsw{K. Narain, M. Sarmadi, and E. Witten, Nucl. Phys. {\bf B279}
(1987) 369.}
\lref\ulf{U. Danielsson and G. Ferretti, hep-th/9610082.} 
\lref\edpaper{E. Witten, Nucl. Phys. {\bf B460} (1996) 335, hep-th/9510135.}
\lref\joenotes{J. Polchinski, S. Chaudhuri, and C. Johnson, 
hep-th/9602052.}
\lref\edjoe{J. Polchinski and E. Witten, Nucl. Phys. {\bf B460} (1996) 525,
hep-th/9510169.} 
\lref\bfss{T. Banks, W. Fischler, S. Shenker, and L. Susskind,
hep-th/9610043.}
\lref\elevend{E. Witten, Nucl. Phys. {\bf B443} (1995) 85, hep-th/9503124.}
\lref\horwitt{P. Horava and E. Witten, Nucl. Phys.
{\bf B460} (1996) 506, hep-th/9510209.} 
\lref\dabharv{A. Dabholkar and J. Harvey, {Phys. Rev. Lett.} {\bf 63}
(1989) 478.}
\lref\Iprime{J. Dai, R. Leigh, and J. Polchinski,
{Mod. Phys. Lett.} {\bf A4} (1989) 2073.} 
\lref\other{
M. Berkooz and M. Douglas, hep-th/9610236\semi 
V. Periwal, hep-th/9611103\semi
L. Susskind, hep-th/9611164\semi 
O. Ganor, S. Ramgoolam, and W. Taylor, hep-th/9611202\semi 
O. Aharony and M. Berkooz, hep-th/9611215\semi
G. Lifschytz and S. Mathur, hep-th/9612087\semi
N. Ishibashi, H. Kawai, Y. Kitazawa, and A.
Tsuchiya, hep-th/9612115\semi
M. Li, hep-th/9612144.}
\lref\mrd{M. Douglas, hep-th/9612126.}
\lref\dkps{M. Douglas, D. Kabat, P. Pouliot, and S. Shenker,
hep-th/9608024.}

\Title{RU-96-114, hep-th/9612162}
{\vbox{\centerline{On Gauge Bosons in the Matrix Model}
        \vskip4pt\centerline{Approach to M Theory}}}
\centerline{Shamit Kachru and Eva Silverstein 
\footnote{$^\dagger$}
{kachru@physics.rutgers.edu, 
evas@physics.rutgers.edu}}
\bigskip\centerline{Department of Physics and Astronomy}
\centerline{Rutgers University}
\centerline{Piscataway, NJ 08855}

\vskip .3in

We discuss the appearance of $E_8\times E_8$ gauge bosons in 
Banks, Fischler, Shenker, and Susskind's zero brane quantum mechanics 
approach to M theory, compactified on the interval $S^1/{\bf Z_2}$.  
The necessary bound states of zero branes
are proven to exist by a straightforward application of T-duality
and heterotic $Spin(32)/{\bf Z_2}$-Type I duality.  
We then study directly the zero brane
Hamiltonian in Type $I^\prime$ theory.  This Hamiltonian 
includes couplings between the
zero branes and background Dirichlet 8 branes localized at the orientifold
planes.
We identify states, localized at the orientifold 
planes, with the requisite gauge boson quantum numbers.
An interesting feature is that $E_8$ gauge symmetry relates bound states of 
different numbers of zero branes.

\Date{12/96} 

\newsec{Introduction}

Recently, an intriguing proposal for the microscopic
description of M theory was given in \bfss.  
Gravitons propagating in eleven dimensions are
bound states of type IIA Dirichlet zero branes 
\elevend.  The Ramond-Ramond 1-form charge carried
by the zero branes maps to momentum in the 11th dimension.
In the infinite momentum frame, the other degrees of freedom
of the IIA theory decouple, leaving a large N matrix quantum mechanics
describing the interactions of the zero branes \refs{\edpaper,\bfss}.
Several tests and explorations of this conjecture have
been performed in \refs{\other,\mrd}.  Previous results about 
maximally supersymmetric quantum mechanics and zero
brane dynamics, which play a role in the conjecture of
\bfss, were obtained in \refs{\maxqm,\ulfI,\KP,\dkps}. 

It is of interest to understand matrix theory in
more general backgrounds with less supersymmetry.
A first step in that direction is to recover the physics
of M theory compactified on $S^1/{\bf Z}_2$, which
is believed to govern the behavior of 
the $E_8\times E_8$ heterotic string at nonzero coupling \horwitt.
In this description, the $E_8\times E_8$ gauge bosons propagate
in ten dimensions on the
ends of the interval $S^1/{\bf Z}_2$.
In this note, we make the simple 
observation that perturbative T-duality symmetries
of string theory together with heterotic/type I duality 
\refs{\elevend,\edjoe} suffice to
prove the existence of the bound states of D0 branes which fill out
$E_8 \times E_8$ gauge multiplets.  
We also formulate conditions for the required states
in the D0 brane quantum mechanics, and describe states
satisfying these conditions for the cases $N=1$ and $N=2$.

One approach to the study of nontrivial backgrounds
in the matrix theory involves starting with type IIA theory
(as in \S9\ of \bfss).
One then constructs M theory as the large N quantum
mechanics of D0 branes (including all associated BPS degrees
of freedom) in the compactified IIA theory.\foot{One
of the most interesting challenges is to understand
how this proposal could be implemented in
situations where BPS states do not exist.}
This is the approach that we take in this paper. 
In section two we review the relevant matrix quantum mechanics
which controls 
the dynamics of zero branes in Type $I^\prime$ theory, as written down
in \ulf.  In section three we present the simple duality argument proving the
existence of the required bound states of zero branes to the D8 branes
and the orientifold planes.  In section four we return to a discussion of
the states directly in the quantum mechanics.
Section five contains some additional remarks.

\newsec{The Type $I^\prime$ Formulation}

We begin with type IIA on the orientifold
$S^1/{\bf Z}_2$, also known as type $I^\prime$ theory
\Iprime.  We will call the $S^1/{\bf Z}_2$ dimension the
tenth dimension, and time the first dimension.
In this theory, which is T-dual to type I string theory,
we consider the locus on Narain moduli space where
8 D8 branes are at each orientifold plane.  
This gives a $Spin(16)\times Spin(16)$ gauge symmetry arising
from open strings with ends on the 8-branes \joenotes.

So far we are in nine dimensions.  As explained in
\elevend, the theory grows a tenth dimension as
$\lambda_{I^\prime}\to\infty$.  
The type $I^\prime$ coupling, $\lambda_{I^\prime}$, determines
the size $R_{11}$ of the eleventh dimension of M theory, as \elevend 
$$R_{11}=\lambda_{I^\prime}^{2/3}.$$  As in \bfss, states
with momentum in the eleventh dimension have nontrivial Ramond-Ramond
1-form charge.  The $Spin(16)\times Spin(16)$ gauge
bosons coming from 8-8 strings do not carry this charge, and
do not have momentum in the eleventh dimension.  

We need to fill out the predicted $E_8\times E_8$ gauge group
which can propagate in ten dimensions (the dimensions 1-9,11
in our notation).  
If we take $R_{11}$ finite, the momentum $p_{11}$ of a
state in the eleventh dimension is given by
\eqn\elevmom{p_{11}={N\over R_{11}}}
where $N$ is the zero brane charge of the state.
As $R_{11}\to \infty$, only states with $N/R_{11}$ finite
have nonzero $p_{11}$. 
The adjoint of $E_8$ decomposes as
the ${\bf 120}+{\bf 128}$ of $Spin(16)$.  So far we have
found a ${\bf 120}$ propagating in the dimensions 1-9.

Some interesting aspects of the quantum mechanics
of this system were presented in \ulf ~-- we follow the notation of
that paper with minor modifications.
The dynamics of N zero branes near the
orientifold plane is (ignoring the 8 D8 branes for now)
described by an $SO(N)$ quantum mechanics
with 8 supersymmetries.  This has
coordinates
$$A^{IJ}_{1,10},~~~X^{IJ}_{2,...,9},~~~{\rm and}~~~x_{2,...,9}$$
where the $A^{IJ}$ are antisymmetric in the $SO(N)$ indices $I$ and $J$
while the $X^{IJ}$ are traceless symmetric representations and
the $x$s are singlets.   The fermionic superpartners are
$$S^{IJ}_{a},~~~S^{IJ}_{\dot a},~~~{\rm and}~~~s_{\dot a}$$ 
with $S_{a}$ in the adjoint, $S_{\dot a}$ in the traceless symmetric,
and $s_{\dot a}$ a singlet.  Here $a$ is an index in the ${\bf 8}_s$
of $SO(8)$ and $\dot a$ is an index in the ${\bf 8}_c$.

Let us now consider the effects of the 8 D8 branes
coincident with the orientifold plane.
We will need degrees of freedom coming from
the 0-8 strings.  Analysis of the worldsheet theory of
the 0-8 strings as in \S4.2\ of \joenew\ reveals that
the Neveu-Schwarz sector has vacuum energy $+{1\over 2}$,
so the only massless states are the Ramond-sector states.
Therefore, we must include fermions
$$\chi^I_{r}$$ 
where $r$ is an $SO(16)$ index.  Thus we have fermions 
in the $(16,N)$
representation of $SO(16)\times SO(N)$, where the $SO(16)$ lives
on the D8 branes.

This system is a modification of that considered in \ulf\
in two respects.  Firstly, we are discussing $SO(N)$ quantum mechanics
instead of $SO(2N)$ (we allow the possibility of unpaired zero branes
which are ``stuck'' to the orientifold plane).  Also, 
we include the vectors of $SO(N)$ arising 
from the 0-8 strings.

The Hamiltonian governing the $SO(N)$ quantum mechanics is
\eqn\hamil{\eqalign{& H ~=~ Tr \biggl\{ \lambda_{I^\prime} ({1\over 2}
P_{i}^{2} - {1\over 2} E_{10}^{2})  
+ {1\over \lambda_{I^\prime}} ({1\over 2}[A_{10},X_i]^2
-{1\over 4} [X_i, X_j]^2 )  \cr
& +{i \over 2} ( - S_{a} [A_{10},S_a] - S_{\dot a} [A_{10},S_{\dot a}] +
2 X_i \sigma^{i}_{a \dot a}\{S_a, S_{\dot a}\} ) \biggr\}\cr
&+\chi_{r}^IA_{1,IJ}\chi_{r}^J+\chi_{r}^IA_{10,IJ}\chi_{r}^J
+\chi^I_{r}B_1^{rs}\chi^I_{s}\cr}}
Here $E_{10}$ and $P_i$ are the momenta conjugate to $A_{10}$ and
$X_i$. As in \ulf, we have ignored
the overall center of mass coordinates $x_{2,\dots,9}$. 
The last term in \hamil\ describes the coupling 
of the spacetime gauge bosons $B_\mu$ (which come from
the 8-8 strings) to the D0 branes.  

There are some subtleties with this
quantum mechanics system, and we will discuss its analysis
in \S4.
For now let us note that
it is not too difficult to motivate the appearance of
bound states in the ${\bf 128}$ (spinor) of
$Spin(16)$ in this quantum mechanics.
Consider a single zero brane in the type $I^\prime$ theory.
It must sit at one of the two orientifold planes (since
the ${\bf Z}_2$ symmetry would require zero branes
to leave the orientifold plane in pairs)\joenotes.  
Quantizing the zero modes of the fermionic open strings between the 0-brane
and the 8 D8 branes gives the ground state the quantum
numbers of the ${\bf 128}$.  
Now suppose there is a bound state consisting of $2k+1$ D0 branes,
i.e. $2k$ zero branes bound to the one stuck
to the orientifold plane.  The $2k$ zero branes will on average
be separated from the orientifold plane, and hence their fermionic
0-8 strings will not contribute degeneracy to the vacuum.

These states and the others in the ${\bf 120}$ that we need to fill out
the ${\bf 248}$ of $E_8$ propagating in $10$d will be
exhibited using T-duality in the next section.
In particular, we will find that bound states of {\it even} numbers
of D0 branes will comprise the ${\bf 120}$ with nonzero $p_{11}$.

\newsec{T-duality and 0 Brane Bound States}

In order to establish the existence of the requisite
bound states, we can use T-duality and heterotic $Spin(32)/{\bf Z_2}$-type I
duality to map the problem to a question about perturbative
states in the heterotic $Spin(32)/{\bf Z_2}$ theory.  
The type $I^\prime$ theory
is T-dual to the type I theory.  The $S^1/{\bf Z}_2$ of radius
$R_{I^\prime}$ in the
type $I^\prime$ theory becomes an $S^1$ of radius
$R_I=1/R_{I^\prime}$ in the type I theory.
D0 branes in the type $I^\prime$
theory map to D1 branes wrapped around the $S^1$
in the type I theory.  These in turn correspond to
wound perturbative heterotic string states.  
The relation between the perturbative heterotic $E_8 \times E_8$
and $Spin(32)/{\bf Z_2}$ theories on a circle was presented in
\ginsparg, and most of what follows is an application of
that result to the problem at hand.

We are interested in BPS bound states of zero branes, which
map to BPS winding states of the $Spin(32)/{\bf Z}_2$ heterotic string
(with Ramond-Ramond charge mapping to winding number).
The left and right moving momentum for heterotic states
with momentum number $m$ and winding number $n$
in the presence of Wilson lines are \nsw
\eqn\pL{p_L=(P+An,{{{m\over 2}-{1\over 4}A^2n-{1\over 2}A\cdot P}
\over R_h}-nR_h)}
\eqn\pR{p_R={{{m\over 2}-{1\over 4}A^2n-{1\over 2}A\cdot P}
\over R_h}+nR_h.}
Here $P$ is a vector in the internal lattice of the 
$Spin(32)/{\bf Z}_2$ theory, and $R_h$ is the radius
of the circle in this theory.
A basis for this lattice is
\eqn\basis{\eqalign{& e_i-e_{i+1}~~~i=1,\dots,15 \cr
&e_{15}+e_{16}\cr
&{1\over 2}\sum_{i=1}^{16}e_i\cr}}
where $e_i,i=1,\dots,16$ are unit vectors in ${\bf R}^{16}$.  
The roots of the form $\pm e_i\pm e_j,~i,j=1,\dots,16$
have $P^2=2$ and give rise to massless gauge bosons.
The last entry corresponds to the spinor weight of $Spin(32)$.
This weight has length squared 4 and gives rise to massive states
in the spinor representation of $Spin(32)$.

We wish to start at the locus in Narain moduli space with
$Spin(16)\times Spin(16)$ gauge symmetry.  This can be achieved
in a standard way by a Wilson line
\eqn\sowils{A=\sum_{i=1}^8{1\over 2} e_i.}
\noindent From \pL\pR, massless perturbative states must satisfy
$P\cdot A=0$.  This leaves the following subset of
the roots of $Spin(32)/{\bf Z}_2$:
\eqn\basisII{\eqalign{& \pm e_i\pm e_{j}~~~i,j=1,\dots,8 \cr
&\pm e_i\pm e_{j}~~~i,j=9,\dots,16.\cr}}
These along with the Cartan generators
form the adjoint of $Spin(16)\times Spin(16)$.

States with nonzero $p_{11}$ map to {\it massive} BPS states
in the heterotic string.
These are states where the right-movers
are in the ground state \dabharv.  
Level matching requires
\eqn\match{p_L^2-p_R^2=2-2N_L.}  
In order to further specify the relevant states, we must
consult the map between heterotic $Spin(32)$ radius $R_h$ and
coupling $\lambda_h$ and those of the type $I^\prime$ theory
$R_{I^\prime}$, $\lambda_{I^\prime}$:
\eqn\map{R_h={1\over{\sqrt{R_{I^\prime}\lambda_{I^\prime}}}}~~~~
\lambda_h={R_{I^\prime}\over{\lambda_{I^\prime}}}.}
The heterotic $E_8\times E_8$ coupling $\lambda_E$ is
related to $R_{I^\prime}$, $\lambda_{I^\prime}$ by
$$\lambda_E={R_{I^\prime}^{3/2}\over{\lambda_{I^\prime}^{1/2}}}.$$
We see that with $\lambda_E$ held fixed,
$R_h\to 0$ as $\lambda_{I^\prime}\to\infty$.  Therefore
from \pL\pR\ we see that states that survive in the limit
of interest must satisfy
\eqn\cond{{{{1\over 2}m-{1\over 4}A\cdot An-{1\over 2}A\cdot P}}=0.}
Then in order to satisfy \match\ one must impose either 
\eqn\rootcond{(P+An)^2=2,~~~~N_L=0}
or
\eqn\rootcondII{(P+An)^2=0,~~~~N_L=1}
The states satisfying \match\cond\rootcondII\ have $n$ even and give
the gravity multiplet.  The vertex operators of the 
$N_L=0$ states are
\eqn\BPSvert{e^{i(p_L+k)X_L}e^{i(p_R+k)X_R}
\zeta_\mu(\partial X_R^\mu +
ik\cdot\psi\psi^\mu).}
Here $k$ is the nine-dimensional spacetime momentum,
and $X^\mu$ and $\psi^\mu$ are worldsheet bosons and fermions.
What are the quantum numbers of the states satisfying
\match\cond\rootcond ?  This depends, as anticipated in
\S2, on whether $n$ is even or odd.  
The condition is that
\eqn\mat{P+n({1\over 2},{1\over 2},{1\over 2},{1\over 2}, {1\over 2},
{1\over 2},{1\over 2},{1\over 2},0,0,0,0,0,0,0,0)}
have norm squared 2.  Let us consider first the case where
$P$ is a linear combination of the first two sets of roots,
$e_i-e_{i+1},i=1,\dots,15$ and $e_{15}+e_{16}$ in \basis.
In other words, take first the case where $P$ has integer entries.
Also take $n=2k+1$ odd.
Then in order to make \mat\ square to 2, $P$ must shift it to
be of the form 
\eqn\spinor{P+(2k+1)A=(\pm{1\over 2},\pm{1\over 2},\pm{1\over 2},\pm{1\over 2},
\pm{1\over 2},\pm{1\over 2},\pm{1\over 2},\pm{1\over 2},0,0,0,0,
0,0,0,0).}
It is easy to check that the possible $P$s lead to 
an even number of + signs in \spinor.  
So we have obtained the spinor (${\bf 128}$,${\bf 1}$) of 
$Spin(16)\times Spin(16)$.  What about the $({\bf 1},{\bf 128})$?
This arises in a similar fashion by including in $P$ shifts
by the spinor weights $\pm {1\over 2}e_1\pm {1\over 2}e_2\pm\cdots\pm 
{1\over 2}e_{16}$
of $Spin(32)$.  

Now if we take $n=2l$ even, and $P$ to have integer entries,
then in order to have \mat\ square to 2 the allowed $P$s must shift it 
to the form 
\eqn\adj{\pm e_i\pm e_j,i,j=1,\dots,8.}
or 
\eqn\adjII{\pm e_i\pm e_j,i,j=9,\dots,16.}
In this way we recover the $({\bf 120},{\bf 1})+({\bf 1},{\bf 120})$ of
$Spin(16)\times Spin(16)$.  

Notice that the constraints \match\cond\rootcond\ leave the gravitons
and the
gauge multiplets that we have just found, i.e. the
$$({\bf 120},{\bf 1})+({\bf 1},{\bf 120})+({\bf 128},{\bf 1})
+({\bf 1},{\bf 128})$$
of $Spin(16)\times Spin(16)$, as the {\it only} surviving states in the limit
$R_{11}\sim \lambda^{2/3}_{I^\prime}\to\infty$ (along with the gravitons,
of course).
These states, by the chain of dualities explained above,
map to bound states of $n$ D0 branes and 8 D8 branes
at the orientifold
planes. We have seen that they have the quantum
numbers of the requisite $E_8\times E_8$ gauge bosons.
An interesting feature of the states is that 
part of the $E_8$ adjoint (the ${\bf 128}$ of 
$Spin(16)$) arises from bound states of odd numbers
of D0 branes, while the rest (the ${\bf 120}$ of $Spin(16)$)
comes from bound states of even numbers of D0 branes.
So gauge symmetry, like Lorentz symmetry, changes the 
D0 brane number of the state.\foot{This has been independently 
noted in a different context in \mrd.} 

\newsec{The D0 brane-D8 brane system revisited}

In this section we will try to understand  
the bound states (whose existence we have established in
\S3) directly in the quantum mechanics
system \hamil.  

Let us first translate the BPS condition for the mass $M$ 
\eqn\bps{M=p_R={{\tilde m}\over{2R_h}}+nR_h}
to the type $I^\prime$ theory.  By making use of the map
\map\ and making a Weyl rescaling of the metric to 
type $I^\prime$ string frame, this becomes
\eqn\bpsIprime{M={\tilde m\over 2} R_{I^\prime}
+{n\over \lambda_{I^\prime}}.}
Here ${1\over 2}\tilde m={1\over 2} m-{1\over 4}A\cdot An
-{1\over 2}A\cdot P$.
So the condition \cond\ that $\tilde m=0$ translates to
the statement that states surviving in the limit 
$\lambda_{I^\prime},R_{I^\prime}\to\infty$ do not
involve open strings stretched across the interval
$S^1/{\bf Z}_2$.  
The Lorentz quantum numbers 
also arise in a simple way from the BPS condition, as 
in the analysis
of the graviton Lorentz quantum numbers in \bfss.  We 
start with 8 supercharges;
a state killed by half the supersymmetries then has
degeneracy $2^4=16$.  This is the correct counting for
a $10d$ $N=1$ vector multiplet.

The challenge is to understand how the
surviving states constitute precisely the gauge bosons
of interest (as well as the graviton).  On the heterotic
side of the duality discussed in \S3, there is an extra
constraint \match.  This arises from the level-matching
constraint $L_0-\bar L_0=0$, where $L_0$, $\bar L_0$ are
the left and right-moving Hamiltonians on the string
worldsheet.  

The D0 brane quantum mechanics system has the following
structure.  If we ignore for now the coupling to
the spacetime gauge fields $B^\mu$, the Hamiltonian \hamil\ takes
the form 
\eqn\hamsplit{H=H_0+\delta H}
where $H_0$ is the part of the Hamiltonian involving
only the 0-0 strings \ulf, and 
\eqn\coup{\delta H=\chi_{r}^IA^1_{IJ}\chi_{r}^J+
\chi_{r}^IA^{10}_{IJ}\chi_{r}^J.}
Here $A^1$ is the $SO(n)$ gauge field, and
$A^{10}$ is its bosonic superpartner which describes the motion of the pair
of zero branes away from the orientifold plane.
The supersymmetry variation of the two terms in $\delta H$ cancel,
so we have
\eqn\deltsup{[Q_a,\delta H]=0.} 
We can take the gauge $A_1=0$ as in \ulf, keeping in
mind that its supersymmetry transformation is involved in
ensuring \deltsup.  As explained in \ulf, there
are operators (supercharges) $Q_a$ which satisfy
\eqn\comq{\{Q_a,Q_b\} =\delta_{ab}H_0.}
Now the fermions $\chi_{r}^I$ do not transform
under supersymmetry, and hence do not appear in the
supercharges, so
\eqn\noncom{\delta H\ne \{Q_a,Q_b\}\zeta^{ab}}
for any constant $\zeta^{ab}$.
This structure is similar to the analogous problem
in the T-dual setup.  There one is interested in
worldsheet properties of the D string.  On its
worldsheet, there are left-moving fermions coming
from the 1-9 strings \edjoe.  The supersymmetry
is right-moving, so that
\eqn\strsusy{\{\bar Q_+,Q_+\}=\bar L_0=H+P.}  
Here $H=L_0+\bar L_0$ is the Hamiltonian, and 
$P=L_0-\bar L_0$ is the momentum.  

We are interested in the quantum numbers of the 
BPS states of this system.
For simplicity, let us begin with $n$ small.
First recall the case $n=1$.  Then, as discussed 
in \S2, the D0 brane is stuck
at the orientifold plane, where 8 D8 branes also lie.
In terms of the formalism \hamil, this means
that the Hamiltonian does not depend on the
fermions $\chi$ at all.
Quantizing the $\chi$ zero modes then gives precisely the 
spinor of $Spin(16)$, the ${\bf 128}$.  

Now let us move on to the case $n=2$.  Here one
has two D0 branes, which can move off the orientifold
plane as a mirror pair.  Let us define
\eqn\chidef{\eqalign{&\chi_{r}
={1\over\sqrt{2}}(\chi^1_{r}+i\chi^2_{r})\cr
&\bar\chi_{r} ={1\over\sqrt{2}}(\chi^1_{r}-i\chi^2_{r})\cr}}
The couplings in $\delta H$
reduce to (in $A_1=0$ gauge)
\eqn\deltnII{\delta H_{n=2}=A^{10}_{12}\sum_{r=1}^{16}\bar\chi_r\chi_r.}  
The canonical anticommutator involving the $\chi$s is
\eqn\com{\{\bar\chi_{r},\chi_{s}\}=\delta_{r,s}.}    
Now take the vacuum  $ \vert 0\rangle$ to
satisfy 
\eqn\vacdef{\chi_r \vert 0\rangle = 0.}
Then 
\eqn\vackill{\delta H \vert 0\rangle =0.}
We expect that an analogue of the GSO projection (a ${\bf Z}_2$ discrete
gauge symmetry \joenotes) will remove states created
by an odd number of $\bar\chi$s.  

The next possibility is the set of states
\eqn\adjoint{\vert {\bf 120}\rangle =\bar\chi_r\bar\chi_s\vert 0\rangle.}
This state transforms in the ${\bf 120}$ of $SO(16)$.  It
corresponds to a configuration of two 0-8 strings.
It satisfies
\eqn\adjen{\delta H\vert{\bf 120}\rangle 
= 2A^{10}_{12}\vert{\bf 120}\rangle.}
So far we have been treating $A^{10}$ as a background field.
We must also take into account the fields 
$X^{IJ}_{2,\dots,9}, S^{IJ}_{\dot a}$ which transform
in the traceless symmetric representation of $SO(2)$.
As pointed out in \ulf, there is a branch classically where $X^{IJ}\ne 0$ and 
$A^{10}=0$.   
We can now begin to develop a microscopic description of how
the gauge bosons are bound to the ends of the world.  From 
\comq, we see that $H_0$ kills the BPS states we
are interested in (which have the quantum numbers of Lorentz
vectors).  This means that in order for a state to
be simultaneously BPS and and an eigenstate of the {\it full}
Hamiltonian, it must also be an eigenstate of $\delta H$.
For our states, which satisfy \adjen, this requires that
\eqn\Akill{A^{10}_{12}\vert {\bf 120}\rangle =0.}
As $X^{IJ}\to\infty$, we can analyze
the system semiclassically and indeed the wavefunction asymptotes
to an eigenvector of $A^{10}_{12}$ with eigenvalue zero.

One puzzling feature of this picture is that there would
seem to be an SO(16)-invariant state, the vacuum
$\vert 0\rangle$, which is also stuck to the wall for
$X^{IJ}\to\infty$.  One difference between this state and
the $\vert {\bf 120}\rangle$ is that $\vert 0\rangle$ 
is not constrained by the analogue of \Akill, since
it satisfies \vackill.  So this state is not required by
BPS to lie at the end of the world.
One possibility is that its wavefunction is generally supported
away from the wall, where the appropriate zero-brane Hamiltonian
has enhanced supersymmetry (it 
becomes the maximally supersymmetric Hamiltonian of \refs{\ulfI,\KP}) and 
these states become gravitons.
 

We have already recovered states with the expected quantum numbers.
What about the states created by more (even) powers of $\bar\chi$?
These states are not protected from decay into
the states we have already found.  It would be interesting
to study these decay processes directly, using the couplings
in the Hamiltonian
between the spacetime gauge field $B^\mu$ and
the quantum-mechanical coordinates $X^{IJ}$, $\chi$, and $A^{10}_{IJ}$.

It is difficult to prove directly the existence of
bound states at threshold with larger $n$.  Assuming
that such bound states exist the analysis given here explains
how the quantum numbers of the ${\bf 128}$ and the ${\bf 120}$ arise.
The indirect derivation using duality in the previous section
ensures that the requisite states exist.



\newsec{Comments}

The existence of a matrix model formulation of the $E_8 \times E_8$
heterotic string theory (M theory on $S^1/{\bf Z_2}$) 
requires the existence of bound states of zero branes to 
D8 branes at orientifold planes in the Type $I^\prime$ theory.
The derivation provided
in \S3 constitutes a proof of their existence assuming only T-duality
and heterotic $Spin(32)/{\bf Z_2}$-type I duality. In this sense proving
the existence of the gauge
bosons with $p_{11}\ne 0$ in the matrix theory is easier than
the analogous problem for the gravitons discussed in \bfss\ (although
Sen has given strong arguments for the existence of the relevant
zero brane bound states in that context as well \Sen). 
This is because the gauge bosons propagate only in ten
dimensions.

In \bfss\ D0 brane bound state scattering was studied and
shown to reproduce low-energy graviton exchange.
This involves exchange of closed string states, which
is an annulus computation in the open string channel.
Scattering of gauge bosons, on the other hand,
involves exchange of {\it open} string states
(the 8-8 strings).  This is simply a tree-level diagram
encoding the scattering of two 0-8 strings (including
the effects of nonzero D0 brane velocity) by exchange
of an 8-8 string.

The heterotic string arises in a straightforward way from
duality as well in this formalism.  The heterotic string
with zero winding number maps to the D string in the type I
theory.  Upon T-duality to type $I^\prime$, this becomes
a D2 brane stretched between the orientifold planes.
It is interesting to think about obtaining the heterotic string
directly in the matrix model by using an analogue of the wrapped
membrane of \bfss.  In the heterotic theory one expects to only
obtain membranes stretched between the two orientifold planes.
Recall that in \bfss\ the membrane extended in the $i,j$
directions is obtained by finding
(infinite) matrices with nonzero $Tr ([X_i,X_j])$.  
This means that, in the notation of \S2,  one expects 
$$Tr([X_i,A_{10}])$$
to be the only nonzero trace in the $N \to \infty$ limit. 
This follows
directly from the symmetry properties of the matrices: The
$X_i$ are symmetric while $A_{10}$ is antisymmetric, so even
in the infinite N limit only commutators with one factor of $A_{10}$
can have nonzero trace.\foot{This was pointed out to us by O. Aharony.}   

The quantum mechanics of the system \hamil\ has a very rich
structure which we have only begun to explore.  There
are at least two features which deserve further exploration:
the change in the zero brane number under $E_8$ gauge
transformations, and the importance of the branch
$X^{IJ}\ne 0$ in producing states bound to the orientifold plane. There are
also interesting resonances \ulf, as in the maximally supersymmetric case 
\refs{\ulfI,\KP,\dkps}.
Decays involving the background spacetime gauge fields would
be interesting to study.  In any case it is clear that applying the
prescription of \bfss\ to the heterotic/type I/type $I^\prime$ string
yields an intriguing microscopic picture of its physics.

\smallskip
\centerline{\bf Acknowledgments}

We are grateful to O. Aharony, T. Banks, W. Fischler, J. Maldacena,
S. Shenker, and L. Susskind
for useful discussions.  This work was supported in part by
DOE DE-FG02-96ER40559.

\listrefs
\end